\documentclass[conference]{ieeetran}
\usepackage{cite}
\usepackage[pdftex]{graphicx}
\usepackage{epstopdf}
\usepackage{amsmath}
\usepackage{cases}
\usepackage[caption=false,font=footnotesize]{subfig}
\usepackage{fixltx2e}
\usepackage{amssymb}
\usepackage{mathtools}
\usepackage{bm}
\usepackage{multirow}
\usepackage{color}
\usepackage{algorithm}
\usepackage{algorithmic}
\usepackage{amsthm}
\usepackage{mathrsfs}

\newcommand{\tabincell}[2]{\renewcommand\arraystretch{0.9}\begin{tabular}{@{}#1@{}}#2\end{tabular}}

\begin{document}


\title{Extended Load Flexibility of Industrial P2H Plants: \\ A Process Constraint-Aware Scheduling Approach}

\author{\centering
  \IEEEauthorblockN{Yiwei~Qiu}
  \IEEEauthorblockA{
  College of Electrical Engineering \\
   Sichuan University \\
   Chengdu, China\\
  ywqiu@scu.edu.cn
  }
  \\
  \IEEEauthorblockN{Yi Zhou}
  \IEEEauthorblockA{
   College of Electrical Engineering \\
   Sichuan University \\
   Chengdu, China \\
   zhouyipower@163.com
  }
  \and
  \IEEEauthorblockN{Buxiang Zhou}
  \IEEEauthorblockA{
  College of Electrical Engineering \\
   Sichuan University \\
   Chengdu, China \\
   hiway\_scu@126.com
  }\\
  \IEEEauthorblockN{Ruomei Qi}
  \IEEEauthorblockA{
  Department of Electrical Engineering \\
  Tsinghua University \\
  Beijing, China \\
  qrm18@mails.tsinghua.edu.cn
  }
  \and
  \IEEEauthorblockN{Tianlei Zang*} 
  \IEEEauthorblockA{
  College of Electrical Engineering \\
   Sichuan University \\
   Chengdu, China \\
   zangtianlei@126.com \\
    }
   \\
  \IEEEauthorblockN{Jin Lin}
  \IEEEauthorblockA{
  Department of Electrical Engineering \\
  Tsinghua University \\
  Beijing, China\\
  linjin@tsinghua.edu.cn
  }
}

\maketitle

\begin{abstract}
  The operational flexibility of industrial power-to-hydrogen (P2H) plants enables admittance of volatile renewable power and provides auxiliary regulatory services for the power grid. Aiming to extend the flexibility of the P2H plant further, this work presents a scheduling method by considering detailed process constraints of the alkaline electrolyzers. Unlike existing works that assume constant load range, the presented scheduling framework fully exploits the dynamic processes of the electrolyzer, including temperature and hydrogen-to-oxygen (HTO) crossover, to improve operational flexibility. Varying energy conversion efficiency under different load levels and temperature is also considered. The scheduling model is solved by proper mathematical transformation as a mixed-integer linear programming (MILP), which determines the on-off-standby states and power levels of different electrolyzers in the P2H plant for daily operation. With experiment-verified constraints, a case study show that compared to the existing scheduling approach, the improved flexibility leads to a 1.627\% profit increase when the P2H plant is directly coupled to the photovoltaic power.
\end{abstract}


\section{Introduction}
\label{sec:intro}

Industrial hydrogen production via water electrolysis has been recognized as promising for renewable electrical power admittance and decarbonization of the chemical engineering and transportation industry \cite{li2021co,klyapovskiy2021optimal}. As an electrical power load, the power-to-hydrogen (P2H) plant can adjust its load level flexibly to accommodate the fluctuating power generation of wind or solar energy \cite{zhang2014wind,li2020capacity} or provide auxiliary regulatory services (peak shaving, frequency regulation, etc.) for the power systems \cite{kopp2017energiepark,el2019hydrogen}. In order to better admit the volatile renewable power and provide auxiliary services, the flexibility of P2H load should be fully exploited.

Water electrolysis is the most common power-to-hydrogen (P2H) conversion method. Mainstream technical routes include alkaline water electrolysis (AEL), proton exchange membrane electrolysis (PEMEL), and solid oxide cell electrolysis (SOCEL) \cite{grigoriev2020current}. Due to relatively high maturity, large capacity, and long lifespan, AEL is preferred by many for industrial hydrogen production \cite{grigoriev2020current,varela2021modeling,straka2021comprehensive}. Hence, this work focus on extending the flexibility of AEL P2H plants.

An alkaline electrolyzer can adjust its hydrogen production rate and power consumption, but its load flexibility is constraint by internal dynamic processes \cite{david2020dynamic,qi2021pressure}. For example, the electrolyzer cannot operate at full power unless warmed up \cite{david2020dynamic} and cannot operate at low power for a long duration due to the accumulation of hydrogen-to-oxygen (HTO) crossover \cite{david2020dynamic,qi2021pressure}. Moreover, the energy conversion efficiency is significantly affected by the temperature and pressure \cite{david2020dynamic}.

In addition, an industrial P2H plant is usually composed of multiple electrolyzers \cite{grigoriev2020current,li2021co,varela2021modeling}. When coupled with renewable power generation, in order to accommodate the intermittent and volatile power supply, the plant operator needs to determine the number of operational electrolyzers and their load levels at different times according to the renewable power forecast \cite{serna2017predictive,varela2021modeling}. Therefore, a plant-level daily scheduling program is also needed.

In traditional P2H plant scheduling methods, the operational constraints of the electrolyzer are set as constant \cite{serna2017predictive,varela2021modeling,uchman2021varying}. However, this can be too conservative. For example, the electrolyzer can operate at low power for a short time without violating the HTO impurity constraint, as explained in Section \ref{sec:hto}. Therefore, to fully extend the load flexibility of the P2H plants, this work aims to present a scheduling approach taking into account these dynamic process constraints that are not considered in existing research. The literature review and contributions of this work are briefed below.

\subsection{Literature Review}
\label{sec:review}

The flexible operation ability of P2H load has been recognized by the community. However, many existing works only considered small-capacity integration of P2H in the renewable power systems or microgrids with only one electrolyzer, and the process constraints were omitted \cite{shams2021machine,fragiacomo2020technical}.

Several latest works investigated the P2H plant scheduling with multiple electrolyzers. For example, rule-based plant scheduling strategies were proposed in \cite{fang2019control,jing2021analysis}, which may not be optimal. Serna et al. \cite{serna2017predictive} proposed a model predictive control (MPC) based scheduling framework for offshore electrolyzers coupled with wind and wave power. Varela et al. \cite{varela2021modeling} developed an MILP-based scheduling model for the P2H plant to consider the on-off operation state of each electrolyzer. 
Uchman et al. \cite{uchman2021varying} used exhaustive search to determine the optimal production schedule of three electrolyzers, which can be hard to scale up. He et al. \cite{he2021hydrogen} considered the UC and capacity limits in P2H plant scheduling, but the process constraints of the electrolyzers were omitted.

In summary, existing works usually assume a constant energy conversion efficiency, and the process constraints are omitted. Instead, the operation of electrolyzer is constrained in a fixed range. These simplifications may adversely affect flexibility. Although electrolyzer control methods considering process dynamics are under investigation \cite{flamm2021electrolyzer,zheng2021optimal,qi2021pressure}, there still lacks a scheduling framework to coordinate multiple electrolyzers to improve plant-level load flexibility.


\begin{figure}[t]
  \centering
  \includegraphics[scale=0.98]{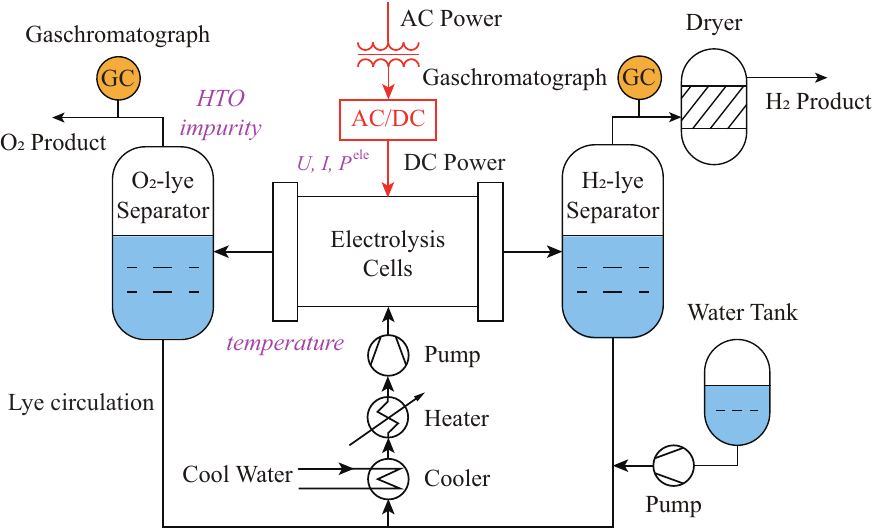}\vspace{-3.pt}
  \caption{Process schematic of an alkaline electrolyzer}
  \label{fig:ale}\vspace{-11.5pt}
\end{figure}

\subsection{Contributions of This Work}

To extend the load flexibility of the industrial P2H plants, a scheduling method considering dynamic process constraints of the alkaline electrolyzer is proposed. The main contributions of this work include:
\begin{enumerate}
    \item A novel scheduling framework for industry P2H plants considering process constraints in the alkaline electrolyzer is first presented.
    \item The proposed P2H plant scheduling model with the nonlinear production function and dynamic temperature and HTO crossover constraints are reformulated and solved as an MILP problem.
    \item Case studies show that the extended flexibility improves total hydrogen production compared to the traditional method with constant constraints when the plant is directly coupled with solar energy.
\end{enumerate}

The remainder is organized as follows. Section \ref{sec:constraint} presents the dynamic process constraints of the alkaline electrolyzer; Section \ref{sec:scheduling} presents the plant scheduling model and solution method, which is verified by case studies in Section \ref{sec:case}.

\section{Modeling Dynamic Process Constraints of an Alkaline Electrolyzer}
\label{sec:constraint}

\subsection{Overview}

The process schematic of an alkaline electrolyzer is shown in Fig. \ref{fig:ale}. To accommodate fluctuating power supply, the electrolyzer switches between three different operational states, namely Production (P), Standby (S), and Idle (I).

In Production, the electrolyzer breaks up water molecules into hydrogen and oxygen using electrical power, the pump keeps lye circulating, the cooler takes away excess heat, and the control system keeps the temperature and pressure in an appropriate interval. The total power consumption is the sum of electrolytic power and the power consumed by the auxiliary equipment.
In Standby, the electrolytic power becomes zero, but the control system keeps working, and the heater keeps the system warm so that it can quickly switch to the Production state.
In Idle, the whole system is turned off.





\subsection{State Switching of Electrolyzers}
\label{sec:state}

\begin{figure}[t]
  \centering
  \includegraphics[scale=0.98]{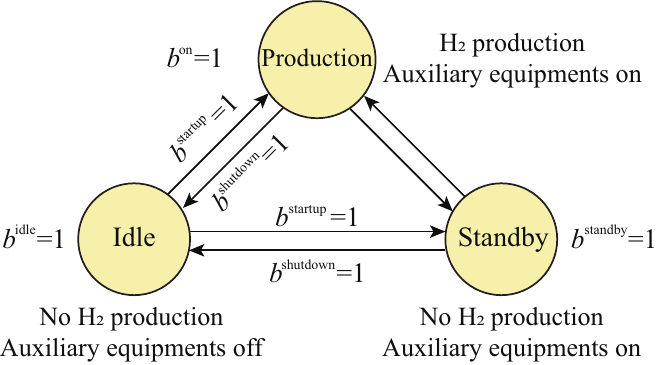}\vspace{-.5pt}
  \caption{Operational states of an electrolyzer}
  \label{fig:states}\vspace{-12pt}
\end{figure}

Following the idea of \cite{varela2021modeling}, three mutually exclusive binary variables $b_{i,k}^{\text{on}}$, $b_{i,k}^{\text{standby}}$, and $b_{i,k}^{\text{idle}}$ are used to represent the three operational states for the $i$th electrolyzer at time $k$, as
\begin{align}
  b_{i,k}^{\text{on}} + b_{i,k}^{\text{standby}} + b_{i,k}^{\text{idle}} = 1. \label{eq:states}
\end{align}

The indicators of startup and shutdown, i.e., switching from Idle to Production or Standby and reversely, are depicted by binary variables $b_{i,k}^{\text{startup}}$ and $b_{i,k}^{\text{shutdown}}$, subjected to
\begin{align}
   b_{i,k}^{\text{on}} +  b_{i,k}^{\text{standby}} +  b_{i,k-1}^{\text{idle}} - 1 &\le b_{i,k}^{\text{startup}},  \label{eq:startup}\\
   b_{i,k}^{\text{idle}} +  b_{i,k-1}^{\text{on}} +  b_{i,k-1}^{\text{standby}} - 1 &\le b_{i,k}^{\text{shutdown}}. \label{eq:shutdown}
\end{align}

Meanwhile, a minimal gap of $N^{(\text{min,idle})}$ time steps between shutdown and startup is required, formulated by
\begin{align}
   \hspace{-1pt} b_{i,k-j}^{\text{idle}} + b_{i,k}^{\text{idle}} - \sum_{l=1}^{j-1}  b_{i,k-j+l}^{\text{idle}} \le 0,\ \forall j = 2, \ldots,  N^{(\text{min,idle})}. \hspace{-1pt} \label{eq:gap}
\end{align}

The schematic of electrolyzer state switching is shown in Fig. \ref{fig:states}, and details can be found in \cite{varela2021modeling}.

\subsection{Hydrogen Production and Power Constraints}
\label{sec:prod}

\begin{figure}[t]
  \centering
  \includegraphics[scale=0.88]{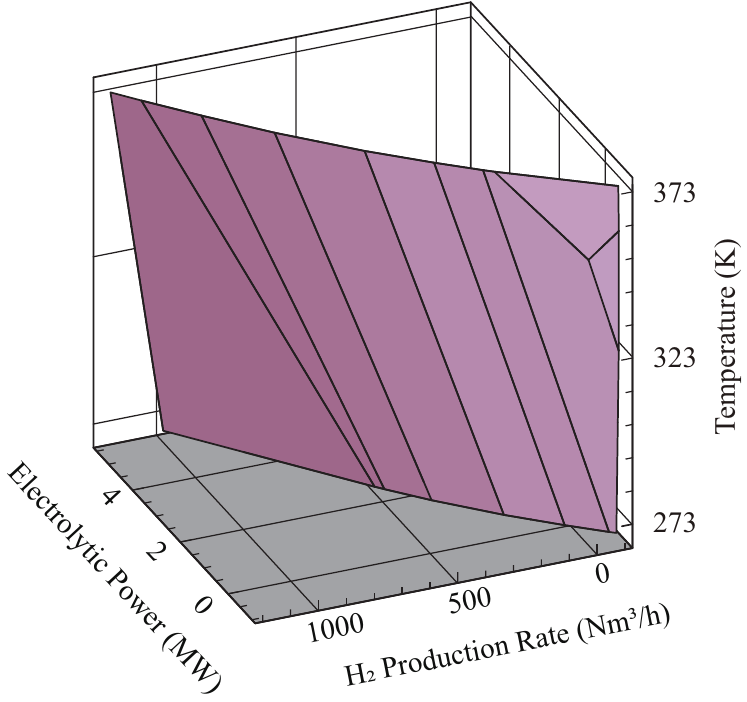}\vspace{-3pt}
  \caption{Piecewise linearized production function of the electrolyzer}
  \label{fig:prod}\vspace{-10pt}
\end{figure}

Supposing the pressure is maintained constant, the hydrogen production rate of an electrolyzer is a concave function of electrolytic power and temperature \cite{li2021co,kopp2017energiepark}, denoted by
\begin{align}
   \dot{n}_{i,k}^{\text{H}_2,\text{prod} } = f(P_{i,k}^{\text{ele}},T_{i,k}). \label{eq:prod}
\end{align}
\noindent
where $\dot{n}_{i,k}^{\text{H}_2,\text{prod} }$ is the $i$th electrolyzer's hydrogen production rate at time $k$; $P_{i,k}^{\text{ele}}$ is the electrolytic power; and $T_{i,k}$ is the temperature of the electrolyzer.

To facilitate modeling the plant scheduling problem as an MILP, the production function (\ref{eq:prod}) is approximated by a polyhedron using the famous double description (DD) algorithm \cite{jones2010polytopic} and then relaxed as a group of inequality constrains, as
\begin{align}
  \dot{n}_{i,k}^{\text{H}_2,\text{prod} } \le \bm{A} P_{i,k}^{\text{ele}} + b_{i,k}^{\text{on}} T_{i,k} \bm{B}  +  b_{i,k}^{\text{on}} \bm{C}, \label{eq:prod}
\end{align}
\noindent
where $\bm{A}$, $\bm{B}$ and $\bm{C}$ are constant coefficient vectors.

According to (\ref{eq:prod}), when the electrolyzer is in Standby or Idle, 
the hydrogen production is zero.
When in Production, because the scheduling objective always maximizes hydrogen production, the operation point will be on the surface of the production function, as shown in Fig. \ref{fig:prod}.


Moreover, to avoid sudden changes in pressure, separator liquid level, or stack gas-liquid ratio that may cause excessive stress, the ramping rate of production is limited, as
\begin{align}
  \underline{r}^{\text{H}_2,\text{prod} } \le \dot{n}_{i,k+1}^{\text{H}_2,\text{prod} }-\dot{n}_{i,k}^{\text{H}_2,\text{prod}} \le \overline{r}^{\text{H}_2,\text{prod}}, \label{eq:ramp}
\end{align}
\noindent
where $\overline{r}^{\text{H}_2,\text{prod}}$ and $\underline{r}^{\text{H}_2,\text{prod}}$ are the upper and lower bounds.

The power consumption of an electrolyzer is the sum of both electrolytic and the balance of plant (BoP) consumption:
\begin{align}
  P_{i,k} = P_{i,k}^{\text{ele}} + (b_{i,k}^{\text{on}} + b_{i,k}^{\text{standby}}) P_{i,k}^{\text{BoP}}. \label{eq:power}
\end{align}

The BoP consumption $P_{i,k}^{\text{BoP}}$ includes
\begin{align}
  P_{i,k}^{\text{BoP}} = P_{i,k}^{\text{heat}}/\eta^{\text{heat}}  + P_{i,k}^{\text{cool}}/\eta^{\text{cool}} + P^{\text{aux}} \label{eq:bop}
\end{align}
\noindent
where $P_{i,k}^{\text{heat}}$ and $P_{i,k}^{\text{cool}}$ are the active heating and cooling power; $\eta^{\text{heat}}$ and $\eta^{\text{cool}}$ are heating and cooling efficiencies; $P^{\text{aux}}$ is the power of auxiliary equipments like the pumps and the control system, which is assumed to be constant here.

When directly coupled to renewable energy sources (RESs) such as photovoltaic power, the total power consumption cannot exceed the available RES power $P_k^{\text{{RES}}}$ at each moment:
\begin{align}
  \sum\nolimits_{i=1}^{N^{\text{ele}}}  P_{i,k} \le P_k^{\text{{RES}}}, \label{eq:totalpower}
\end{align}
\noindent
where $N^{\text{ele}}$ is the number of electrolyzers in the plant.

\subsection{Temperature Dynamic Model and Constraints}
\label{sec:temp}

\begin{figure}[t]
  \centering
  \includegraphics[scale=0.95]{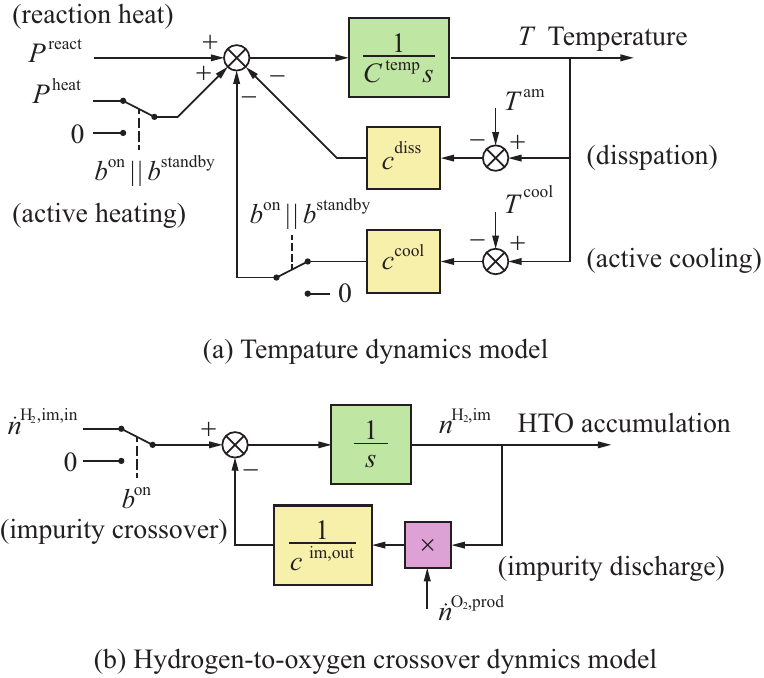}\vspace{-3pt}
  \caption{Diagrams of the temperature dynamic model and HTO crossover dynamic model of an alkaline electrolyzer.}
  \label{fig:temp}\vspace{-12pt}
\end{figure}

Temperature has a big impact on the efficiency and loading range of the electrolyzer \cite{grigoriev2020current,flamm2021electrolyzer,zheng2021optimal}.
Therefore, different from previous works that omitted the temperature dynamics \cite{serna2017predictive,varela2021modeling,uchman2021varying}, this paper takes it into account in scheduling.

The first-order temperature model \cite{flamm2021electrolyzer,zheng2021optimal} is adopted in this work, illustrated in Fig. \ref{fig:temp}(a) and expressed as
\begin{align}\hspace{-3pt}
  T_{i,k+1}  = T_{i,k}   +  h
   &
    \frac{P_{i,k}^{\text{react}} -c^{\text{diss}} (T_{i,k} - T^{\text{am}})  - P_{i,k}^{\text{cool}} + P_{i,k}^{\text{heat}} } {C_i^\text{temp}} \hspace{-3pt} \label{eq:temp}
\end{align}
\noindent
where $C^{\text{temp}}$ is system heat capacity; $c^{\text{diss}}$ is the conductivity of heat dissipation; $T^{\text{am}}$ is the ambient temperature; $h$ is the step length of scheduling; $P_{i,k}^{\text{heat}}$ is the external heating
\begin{align}
  0\le P_{i,k}^{\text{heat}} \le (b_{i,k}^{\text{on}} + b_{i,k}^{\text{standby}}) \overline{P}_{i,k}^{\text{heat}}, \label{eq:heat}
\end{align}
\noindent
where $\overline{P}_{i,k}^{\text{heat}}$ is the upper limit of heating power;
$P_{i,k}^{\text{cool}}$ is the active cooling power, which satisfies
\begin{align}
  0 \le P_{i,k}^{\text{cool}} \le (b_{i,k}^{\text{on}} + b_{i,k}^{\text{standby}}) c^{\text{cool}} (T_{i,k} - T^{\text{cool}}), \label{eq:cool}
\end{align}
\noindent
where $T^{\text{cool}}$ is the coolant temperature; and $P_{i,k}^{\text{react}}$ is the electrolytic heating power, approximated by a second-order function of stack current and temperature, as
\begin{align}
  P_{i,k}^{\text{react}} & = N^{\text{cell}} I_{i,k} \big(U_{i,k}^{\text{cell}} - U^{\text{th}}  \big)  \nonumber \\
    & \approx N^{\text{cell}}  \big( a_0 I_{i,k} + a_1 I_{i,k} T_{i,k} + a_2 I^2_{i,k} -  U^{\text{th}} I_{i,k} \big), \label{eq:heatreact}
\end{align}
\noindent
where $N^{\text{cell}}$ is the number of electrolysis cells in the stack; $U^{\text{th}}= 1.48$ V is the thermal neutral voltage; $a_0$, $a_1$ and $a_2$ are constant coefficients; $I_{i,k}$ is the stack current, which satisfies
\begin{align}
    \dot{n}_{i,k+1}^{\text{H}_2,\text{prod}} = N^{\text{cell}} \eta I_{i,k} / (2F), \label{eq:current}
\end{align}
\noindent
where $\eta$ is the Faraday efficiency of the electrolyzer; and  $F = 96 485.3$ C/mol is the Faraday constant.

Although higher temperature means higher energy conversion efficiency, the stack temperature should stay below a limit to avoid damaging the diaphragm, as
\begin{align}
  T_{i,k} \le \overline{T},
\end{align}
\noindent
where $\overline{T}$ is the upper limit of stack temperature, set as $373$ K in this work; and the electrolytic voltage should not exceed a safety margin (2.1 V here) to avoid damaging electrode microstructure, especially when the temperature is low, as
\begin{align}
  U_{i,k}^{\text{cell}} \big( \approx a_0 + a_1 T_{i,k} + a_2 I_{i,k} \big) \le 2.1(V).
\end{align}

\subsection{Hydrogen-to-Oxygen Crossover Dynamics and Constraints}
\label{sec:hto}

Hydrogen to oxygen (HTO) impurity crossover may cause a flammable mixture. For safety, the electrolyzer will shut down when the hydrogen impurity in the oxygen product reaches 2\% in volume, about half of the explosion limit \cite{straka2021comprehensive,qi2021pressure}.

Due to the HTO impurity accumulation rate being higher at a low load level, the electrolyzers usually have a minimal steady-state load level between 10\% and 40\% \cite{buttler2018current}. 
Existing works on P2H plant scheduling usually assume such a minimal loading constraint \cite{serna2017predictive,varela2021modeling,uchman2021varying,he2021hydrogen}.

However, because HTO accumulation is a dynamic process, operating at a lower power level temporarily without violating the 2\% constraint is possible, which may improve the load flexibility of the electrolyzer. Therefore, this work takes the HTO crossover dynamics into account.

A simplified dynamic model of HTO accumulation is illustrated in Fig. \ref{fig:temp}(b) and can be represented \cite{qi2021pressure} by
\begin{align}
    n_{i,k+1}^{\text{H}_2,\text{im}} =  n_{i,k}^{\text{H}_2,\text{im}} + h \bigg[ b_{i,k}^{\text{on}} \dot{n}^{\text{H}_2,\text{im},\text{in}} - \frac{ \dot{n}_{i,k}^{\text{O}_2,\text{prod}} {n}_{i,k}^{\text{H}_2,\text{im}}}{ c^{\text{im,out}}}  \bigg], \label{eq:hto}
\end{align}
\noindent
where $\dot{n}^{\text{H}_2,\text{im},\text{in}}$ is the impurity crossover flow rate, which can assumed to be constant at constant system pressure, and is non-zero in only Production state; $\dot{n}_{i,k}^{\text{O}_2,\text{prod}}$ is the production rate of oxygen, equal to $\dot{n}_{i,k}^{\text{H}_2,\text{prod}}/2$; $c^{\text{im,out}}$ is the impurity discharge constant, see details in \cite{qi2021pressure}.

The HTO impurity constraint is expressed as
\begin{align}
    {n_{i,k}^{\text{H}_2,\text{im}} / n_{i,k}^{\text{O}_2,\text{prod}} } \le 2\%. \label{eq:conshto}
\end{align}

\begin{figure}[t]
  \centering
  \includegraphics[scale=0.85]{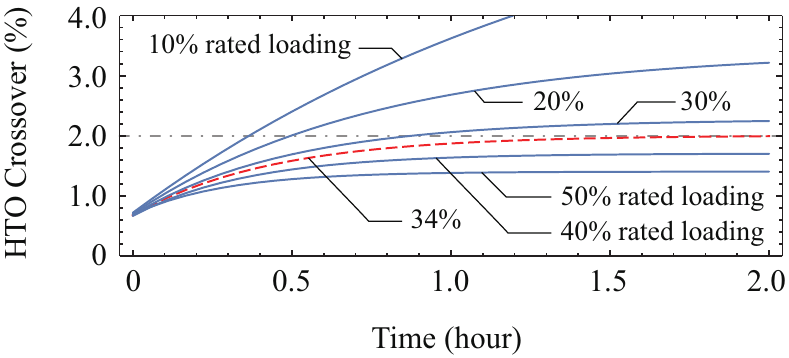}\vspace{-3pt}
  \caption{HTO crossover simulation of the alkaline electrolyzer with parameters given in Table \ref{tab:para} at different steady-state loading levels}
  \label{fig:hto}\vspace{-10pt}
\end{figure}

For easy understanding, Fig. \ref{fig:hto} shows the results of HTO impurity crossover simulation under different steady-state loading levels with electrolyzer parameters given in Table \ref{tab:para}. We can see the HTO impurity accumulates at different speeds under different load levels. If the loading stays lower than 34\%, the HTO will eventually exceed the 2\% safety limit.

The dynamic temperature and HTO crossover models introduced in this section are validated by experiments; see the detailed result in the Appendix of this paper.

\section{The P2H Plant Scheduling Model}
\label{sec:scheduling}

\subsection{Objective}
\label{sec:obj}

The objective of P2H plant scheduling is as maximizing the profit, i.e., the revenue from the sale of hydrogen minus electricity and startup costs, formulated by
\begin{align}
    \hspace{-6pt} J = \sum_{k=1}^{N^{\text{h}}} \sum_{i=1}^{N^{\text{ele}}} \left[ c^{\text{H}_{\text{2}}}  \dot{n}_{i,k}^{\text{H}_2,\text{prod} } - c^{\text{power}}_k P_{i,k} + c^{\text{startup}} b_{i,k}^{\text{startup}} \right] \hspace{-6pt} \label{eq:obj}
\end{align}
\noindent
where $N^{\text{h}}$ is the scheduling horizon; $c^{\text{H}_{\text{2}}}$ is the hydrogen price; $c^{\text{power}}_k$ is the price of electric power at time $k$; and $c^{\text{startup}}$ is the startup cost, which is used to characterize the depreciation expense of the electrolyzer.

\subsection{Reformulation of Process Constraints}
\label{sec:overall}

Some dynamic process constraints introduced in Section \ref{sec:constraint} include bilinear terms. This causes nonlinear formulation of the plant scheduling problem and makes it hard to solve. Therefore, we reformulate them into linear ones.

The bilinear terms are categorized as two types, i.e., the product of a real variable and a binary variable in (\ref{eq:prod}), (\ref{eq:power}), (\ref{eq:cool}), and (\ref{eq:hto}), and the product of two real variables in (\ref{eq:heatreact}) and (\ref{eq:hto}).
For the first type, for example $b_{i,k}^{\text{on}} T_{i,k}$ in (\ref{eq:prod}), it is linearized by the standard big-M method.  

For the second type, i.e., the product of two real variables, for example $I_{i,k} T_{i,k}$ in (\ref{eq:heatreact}), we can discretize it as
  $I_{i,k} \approx \sum_{j=1}\nolimits^{N^\text{d}} 2^j  \beta^I_{i,k,j}  \Delta I,$
where $\beta^I_{i,k,j}$ is a binary variable and $\Delta I$ is the step length. Then, we can reformulate the product as
\begin{align}
  I_{i,k} T_{i,k} & = \sum_{j=1}\nolimits^{N^\text{d}} 2^j \delta_{i,k,j}^{I,t}  \Delta I, \label{eq:m5} \\
  T_{i,k} - M (1 - \beta^I_{i,k,j} ) & \le  \delta_{i,k,j}^{I,t} \le T_{i,k} + M (1 - \beta^I_{i,k,j} ), \\
   - M \beta^I_{i,k,j} ) & \le  \delta_{i,k,j}^{I,t} \le  M  \beta^I_{i,k,j}. \label{eq:m7}
\end{align}

So far, 
the dynamic process constraints are transformed into linear ones. We can then formulate the overall plant scheduling problem as a mixed-integer linear programming (MILP).

\subsection{Overall Formulation}
\label{sec:overall}

The overall plant scheduling model is summarized as
\begin{align}
   \max_{\bm{u}}\ &(\ref{eq:obj})  \
   \text{subject to}\ (\ref{eq:states})\text{--}(\ref{eq:gap})\ \text{and}\ (\ref{eq:prod})\text{--}(\ref{eq:conshto}),  \label{eq:overallcons}
\end{align}
\noindent
where in the process constraints (\ref{eq:prod}), (\ref{eq:power}), (\ref{eq:cool}), (\ref{eq:heatreact}), and (\ref{eq:hto}) the bilinear terms are replaced by (\ref{eq:m5})--(\ref{eq:m7}).

The control variables $\bm{u}$ include the states indicators, i.e., $b_{i,k}^{\text{on}}$, $b_{i,k}^{\text{standby}}$, $b_{i,k}^{\text{idle}}$, $b_{i,k}^{\text{startup}}$, and $b_{i,k}^{\text{shutdown}}$, and heating and cooling powers $P_{i,k}^{\text{ele}}$, $P_{i,k}^{\text{heat}}$, and $P_{i,k}^{\text{cool}}$ for time $k=1,\ldots,N^h$. They are implemented to the electrolyzers in operation.

The plant scheduling problem (\ref{eq:overallcons}) is an MILP, which can be solved easily with commercial solvers.
Further, notice that the renewable power supply may deviate from the forecast in online operation. In this case, we can implement it in a receding-horizon manner to alleviate the impact of the forecast error. Due to the space limit, we will not go further.

\section{Case Study}
\label{sec:case}

\subsection{Case Settings}
\label{sec:setting}

We assume a P2H plant composed of 6 alkaline electrolyzers, each rated 5 MW, a total rating of 30 MW, connected to photovoltaic power.
The parameters used in the case study are given in Table \ref{tab:para}. Moreover, we assume there is a bilateral contract with the photovoltaic plant. Therefore, the electricity price is set as constant. The platform for modeling the scheduling problem and simulation is \emph{Wolfram Mathematica 12.3}, and the solver for MILP employed is \emph{Gurobi 9.5.0}.

\subsection{Result of the Proposed P2H Plant Scheduling}
\label{sec:result}

\begin{table}[tb]\scriptsize
  \renewcommand{\arraystretch}{1.3}
  \caption{Parameters of the Case Study}\vspace{-4pt}
  \label{tab:para}
  \centering
  \begin{tabular}{cc}
  \hline \hline
  Parameter                                             & Value \\  \hline
  Scheduling horizon and step length $N^h$, $h$         & 96, 15 min  \\
  Electrolyzer number, rated loading $N^{\text{ele}}$, $P^{\text{ele},\text{rated}}$                  & 6, 5 MW    \\
  Hydrogen and electricity price $c^{\text{H}_{\text{2}}}$, $c^{\text{power}}$      & 0.38 \$/Nm$^3$, 34.7 \$/MWh            \\
  Startup cost $c^{\text{startup}}$      & 280 \$            \\
  Production ramp up/down rate $\overline{r}^{\text{H}_2,\text{prod} } $, $\underline{r}^{\text{H}_2,\text{prod} }$    & 1600, $-$4800 Nm$^3$/h$^2$         \\
  Production function $f(\cdot)$           & See Fig. \ref{fig:prod}                   \\
  Ambient, coolant, limit temperature $T^{\text{cool}}$, $T^{\text{cool}}$, $\overline{T}$  & 298, 278, 373 K \\
  Heat capacity and dissipation constant $C^{\text{temp}}$, $c^{\text{diss}}$        & 447.2 MJ/K, 0.033 MW/K   \\
  Impurity crossover flow rate $\dot{n}^{\text{H}_2,\text{im},\text{in}}$ &0.003182 mol/s  \\
  Impurity discharge constant $c^{\text{im,out}}$ & 5.68$\times$10$^5$ mol$^{-1}$ \\
  \hline \hline
  \end{tabular}\vspace{-2pt}
\end{table}

\subsubsection{Basic case}
\begin{figure}[t]
  \centering
  \includegraphics[scale=0.85]{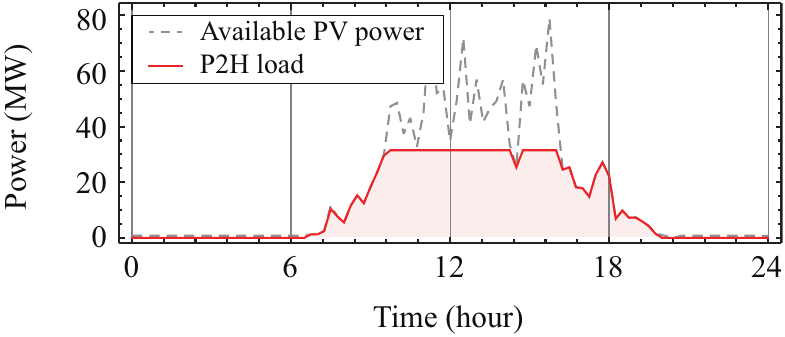}\vspace{-3pt}
  \caption{Available power supply by the photovoltaic station \cite{qiu2020stochastic} and the power consumption of the P2H plant under the proposed scheduling method}
  \label{fig:power}
  \vspace{10pt}
  \centering
  \includegraphics[scale=0.97]{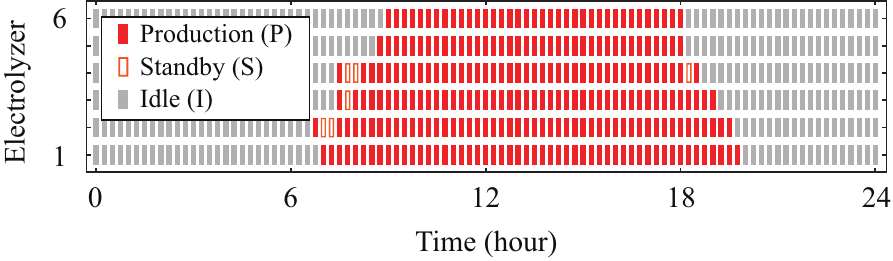}\vspace{-3pt}
  \caption{States of the six electrolyzers under the proposed scheduling method}
  \label{fig:production}
  \vspace{10pt}
  \centering
  \includegraphics[scale=0.85]{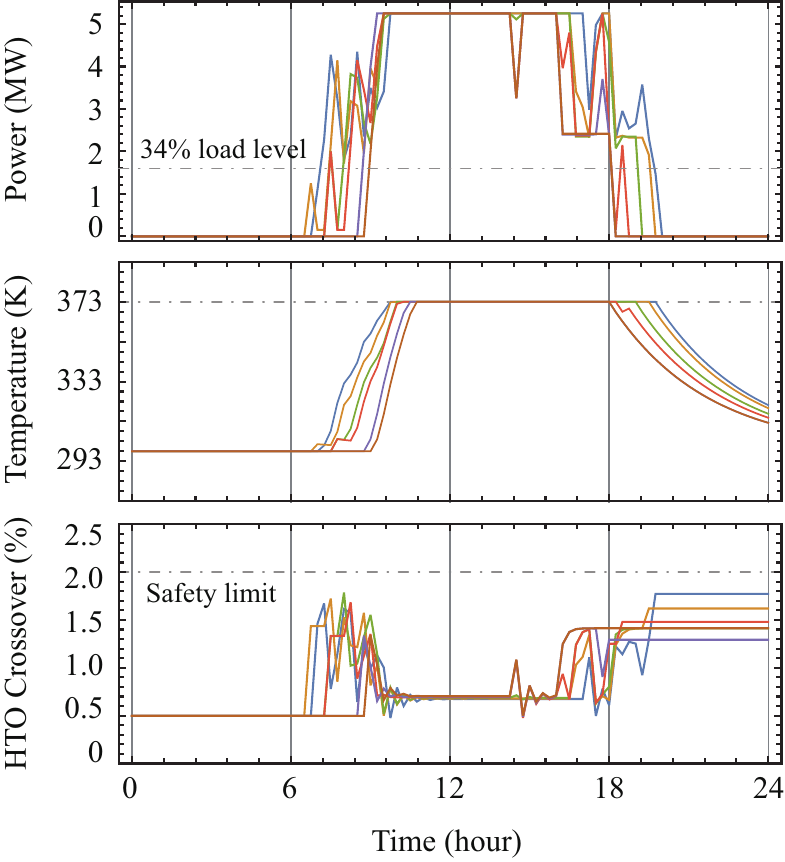}\vspace{-3pt}
  \caption{Loading power, temperature, and HTO impurity of the electrolyzers}
  \label{fig:process}\vspace{-12pt}
\end{figure}

Given the scenario of the power supply based on the data of a photovoltaic plant in Sichuan Province, China, as shown in Fig. \ref{fig:power}, the proposed scheduling method calculates the optimal operational states and power commands for all electrolyzers, respectively shown in Figs. \ref{fig:production} and \ref{fig:process}. Then by simulation of the models from Section \ref{sec:constraint}, the temperature and HTO impurity of each electrolyzer are given in Fig. \ref{fig:process}. 

As seen from Fig. \ref{fig:production}, following the sunrise around 7 am, the electrolyzers startup successively. Then, due to the temporary drop of the solar power, three of them switch to Standby for 15 to 30 minutes before all electrolyzers start production. The power of each electrolyzer increases gradually as the temperature increases to avoid the cell voltage becoming too high. Before reaching the full loading level, the HTO impurity first increases due to a temporarily low loading level and then decreases. After sunset, the electrolyzers switch off and cool down due to heat dissipation.

For the most time, the electrolyzers operate at the upper-temperature limit to maximize production. Meanwhile, the HTO impurity is relatively low due to the high loading, as explained in Fig. \ref{fig:hto}. Note that although the minimal steady-state loading is 34\% and is set as the lower power limit in the literature \cite{serna2017predictive,varela2021modeling,uchman2021varying}, the proposed scheduling method enables the electrolyzers to operate at a lower power level, which extends the flexibility when the power supply is low.

Finally, we compare the proposed process constraint-aware scheduling approach to the existing one \cite{varela2021modeling}, with the result shown in Table \ref{tab:comparison}. We can see that the improved flexibility leads to a 0.825\% increase of the total hydrogen production and a 1.627\% profit increase.

\subsubsection{Comparison to the existing P2H plant scheduling method under various scenarios}
Moreover, we compared the proposed and the existing plant scheduling method using the photovoltaic power of 30 different days. The simulation result shows that the proposed method achieves a 0.839\% hydrogen production increase and a 1.638\% profit increase on average. This result further confirms the improvement of this work.

\begin{table}[tb]\scriptsize
  \renewcommand{\arraystretch}{1.55}
  \caption{Comparison of The Total Hydrogen Production and Profit Between the Existing and the Proposed Scheduling Methods}\vspace{-4pt}
  \label{tab:comparison}
  \centering
  \begin{tabular}{ccc}
  \hline \hline
  Scheduling method                               & H2 Production   & Profit \\ \hline
  \vspace{4pt}Tradition (w/o process constraints)         & $54648.6$ Nm$^3$         & $8980.54$ $\$$ \\
  \vspace{2pt}Proposed (with process constraints)      & \tabincell{c}{$55099.5$ Nm$^3$\vspace{1.5pt} \\ ($+0.825\%$) }          & \tabincell{c}{$9126.52$ $\$$\vspace{1.5pt} \\($+1.627\%$)} \\
  \hline \hline
  \end{tabular}\vspace{-9pt}
\end{table}

\section{Conclusions}

This paper first incorporates the dynamic process constraints of the electrolyzers into the P2H plant scheduling framework. Simulation shows that the proposed method extends the loading flexibility of the P2H plant, which leads to an increase in hydrogen production and profit when the plant is directly coupled to volatile renewable power sources.

Yet, this work has not considered the uncertainty of the power supply. To address this, we may employ a receding-horizon scheduling approach combing the latest stochastic optimization methods to alleviate the impact uncertainty.

Moreover, the hydrogen produced by an industrial P2H plant is generally consumed by synthetic chemical processes, e.g., ammonia or methanol synthesis. However, their requirements on hydrogen supply are not considered in this work. Further, considering the dynamic process constraints of the downstream hydrogen consumers could also be future works.

\appendices
\section{Experiment Validation of Dynamic Process Model}

\begin{figure}[t]
  \centering
  \includegraphics[width=2.6in]{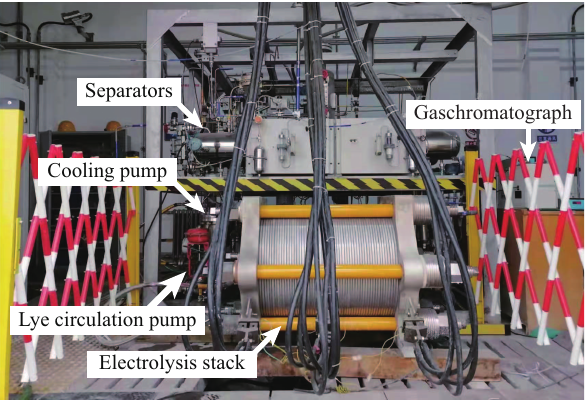}\vspace{-1pt}
  \caption{The CNDQ5/3.2 alkaline electrolyzer used for model validation}
  \label{fig:photo}
  \vspace{12pt}
  \includegraphics[scale=0.80]{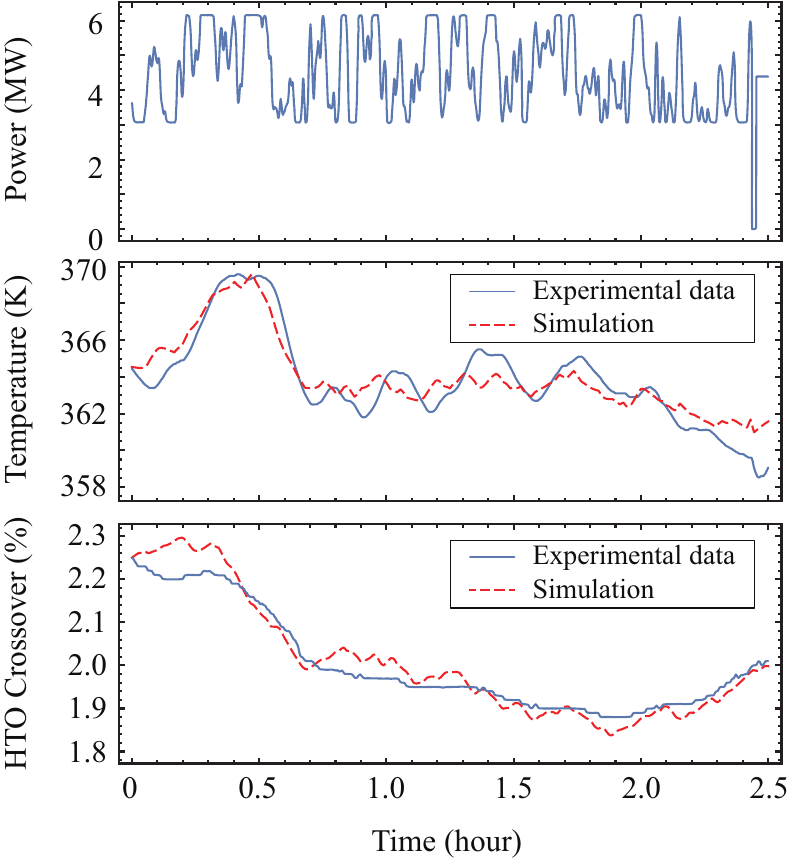}\vspace{-4pt}
  \caption{Experimental data of temperature and HTO impurity crossover compared with simulations of the dynamic models presented in Section \ref{sec:constraint}}
  \label{fig:validation} \vspace{-2pt}
\end{figure}

The dynamic temperature and HTO crossover models presented in Sections \ref{sec:temp} and \ref{sec:hto} are verified by experiments on a CNDQ5/3.2 alkaline electrolyzer manufactured by the Purification Equipment Research Institute of China Shipbuilding Industry Corporation (CSIC), as shown as Fig. \ref{fig:photo}. Its rated hydrogen production rate is $5$ Nm$^3$/h.

We use a rescaled PJM RegD frequency regulation signal on Dec. 1, 2019 \cite{pjmsginal} as the electrolytic power reference. The observed temperature and HTO impurity and the simulation results of the models (\ref{eq:temp}) and (\ref{eq:hto}) are compared in Fig. \ref{fig:validation}. We can see that the simulation fits the experimental data quite well. Therefore, these models are validated.

\section*{Acknowledgement}

Financial supports from National Key R\&D Program of China (2021YFB4000500) and National Natural Science Foundation of China (51907099, 51907097) are acknowledged.

\bibliographystyle{IEEEtran}
\bibliography{IEEEabrv,AWE-SCHD}

\end{document}